
\input harvmac.tex

\def\inbar{\,\vrule height1.5ex width.4pt depth0pt}
\def\IB{\relax{\rm I\kern-.18em B}}
\def\IC{\relax\hbox{$\inbar\kern-.3em{\rm C}$}}
\def\ID{\relax{\rm I\kern-.18em D}}
\def\IE{\relax{\rm I\kern-.18em E}}
\def\IF{\relax{\rm I\kern-.18em F}}
\def\IG{\relax\hbox{$\inbar\kern-.3em{\rm G}$}}
\def\IH{\relax{\rm I\kern-.18em H}}
\def\II{\relax{\rm I\kern-.18em I}}
\def\IK{\relax{\rm I\kern-.18em K}}
\def\IL{\relax{\rm I\kern-.18em L}}
\def\IM{\relax{\rm I\kern-.18em M}}
\def\IN{\relax{\rm I\kern-.18em N}}
\def\IO{\relax\hbox{$\inbar\kern-.3em{\rm O}$}}
\def\IP{\relax{\rm I\kern-.18em P}}
\def\IQ{\relax\hbox{$\inbar\kern-.3em{\rm Q}$}}
\def\IR{\relax{\rm I\kern-.18em R}}
\font\cmss=cmss10 \font\cmsss=cmss10 at 7pt
\def\IZ{\relax\ifmmode\mathchoice
{\hbox{\cmss Z\kern-.4em Z}}{\hbox{\cmss Z\kern-.4em Z}}
{\lower.9pt\hbox{\cmsss Z\kern-.4em Z}}
{\lower1.2pt\hbox{\cmsss Z\kern-.4em Z}}\else{\cmss Z\kern-.4em Z}\fi}
\def\IGa{\relax\hbox{${\rm I}\kern-.18em\Gamma$}}
\def\IPi{\relax\hbox{${\rm I}\kern-.18em\Pi$}}
\def\ITh{\relax\hbox{$\inbar\kern-.3em\Theta$}}
\def\IOm{\relax\hbox{$\inbar\kern-3.00pt\Omega$}}

\magnification1200

\def\CP {{\cal P }}

\def\CO {{\cal O}}
\def\CZ {{\cal Z}}
\def\p {\partial}
\def\CS {{\cal S}}

\def\inbar{\,\vrule height1.5ex width.4pt depth0pt}
\def\IB{\relax{\rm I\kern-.18em B}}
\def\IC{\relax\hbox{$\inbar\kern-.3em{\rm C}$}}
\def\IP{\relax{\rm I\kern-.18em P}}
\def\IR{\relax{\rm I\kern-.18em R}}

\def\log {{\rm log}}

\Title{\vbox{\baselineskip12pt\hbox{YCTP-P1-92}
\hbox{hepth@xxx/9203061}}}
{\vbox{\centerline{Gravitational Phase Transitions}
\centerline{and the}
\centerline{Sine-Gordon Model}}}

\centerline{Gregory Moore}
\centerline{Department of Physics}
\centerline{Yale University}
\centerline{New Haven, CT 06511-8167}
\bigskip
\noindent
We consider the Sine-Gordon model coupled to 2D
gravity. We find a nonperturbative
expression for the partition function as
a function of the cosmological constant, the
SG mass and the SG coupling constant.
At genus zero, the partition function exhibits
singularities which are interpreted as signals of
phase transitions. A semiclassical picture of one of
these transitions is proposed. According
to this picture, a phase in which the Sine-Gordon
field and the geometry are frozen melts into
another phase in which the fields and geometry
become dynamical.

\Date{March 22, 1992}
\noblackbox

\newsec{Introduction}

2D string theory has thus far been most thoroughly studied
in a specific background, defined by the standard gaussian model
coupled to c=25 Liouville theory. An important open problem is
to obtain an equally complete description of strings moving
in other 2D backgrounds.

One approach to this problem is based on
perturbing  the free action for two uncompactified real fields $\phi,X$:
\eqn\liugaus{
S_{\rm Liouville}+S_{\rm Gaussian}=
\int d^2z \sqrt{\hat g}
\biggl({1\over 8 \pi} (\hat\nabla \phi)^2
+{Q\over 8 \pi}\phi R(\hat g)\biggr)+\int d^2z \sqrt{\hat g}
{1\over 8 \pi}(\hat\nabla X)^2 }
by an operator $\sum e^{\alpha_i \phi}\CO_i$, where
$\CO_i$ are operators in the c=1 gaussian model.
In this paper we study the example of
the Euclidean Sine-Gordon model coupled to 2D gravity:
\eqn\pertlag{\eqalign{
S=&\int d^2z \sqrt{\hat g}\biggl[{1\over 8 \pi} (\hat\nabla \phi)^2
+{\mu\over 8 \pi \gamma^2}e^{\gamma \phi}+{Q\over
8 \pi}\phi R(\hat g)\biggr]\cr
&+\int d^2 z \sqrt{\hat g}\biggl[
{1\over 8 \pi} (\hat\nabla X)^2+ m e^{\xi \phi}
\cos (p X/\sqrt{2})
\biggr]\cr}
}
Here $\hat g$ is some background metric.
In flat space the Sine-Gordon model is not
conformal for $m\not=0$.
Coupling the Sine-Gordon model to gravity produces
a nontrivial $c=26$ conformal field theory.
General covariance (and hence conformal
invariance) is maintained in the quantum theory for
$\gamma=\sqrt{2}$, $Q=\sqrt{8}$, $\xi=\gamma(1-|p|/2)$.
We generally use the
notation, conventions, (and insights) of
\ref\nati{N. Seiberg,
Notes on Quantum Liouville Theory and Quantum Gravity,''
In {\it Common Trends in Mathematics and Quantum Field Theory}
Proc. of the 1990 Yukawa International Seminar,
Prog. Theor. Phys. Suppl {\bf 102}}.
The $c=1$ model in 2D gravity is reviewed in
\ref\klebrev{I.R. Klebanov ``String Theory in Two Dimensions,''
Princeton preprint, hepth@xxx/ 9108019.}.

Correlation functions in the theory
\pertlag\  will be defined by ``conformal perturbation
theory.'' That is, introducing the vertex operator
\eqn\vert{
V_p\equiv \int d^2z \sqrt{\hat g} e^{\xi  \phi}e^{ip X/\sqrt{2}}
}
we define
correlation functions at $m\not=0$ by the series
\eqn\backdp{
\langle \prod V_{q_i}e^{\half m (V_p + V_{-p})}\rangle
\equiv\sum_{n_1,n_2\geq 0}{m^{n_1+n_2}\over 2^{n_1+n_2} n_1! n_2!}
\langle \prod V_{q_i} (V_p)^{n_1} (V_{-p})^{n_2}\rangle
}
The coefficients in the expansion \backdp\
are calculated in the standard background with $m=0$
but $\mu\not=0$.
Recent results on the c=1 matrix model have yielded a
complete set of formulae for c=1 correlators
\ref\mpr{G. Moore, R. Plesser, S. Ramgoolam, ``Exact S-Matrix for
2D String Theory,'' Yale preprint P35-91, hepth@xxx/ 9111035}.
In this paper we use these formulae to
learn about the theory \pertlag. Our main result is the
phase diagram shown in figs.2 and 3, and described in section
four. Some physical interpretations of this diagram
are proposed in section five.

\newsec{Flows and phase transitions in $2D$ gravity}

In this section we review some (well-known) aspects of coupling
constant flows in the $c<1$ models coupled to
2D gravity. The discussion is meant to put the phase transitions
discussed in sections four and five into perspective.

The continuum approach to the $c<1$ models
begins with a ``$(p,q)$ theory''
which is a tensor product of a Liouville theory with the
minimal model $M_{p,q}$. Perturbations around
this theory are defined by an action
\eqn\minpert{S=\CS_{p,q}+\sum_{r,k}
\tau_{r,k}\CO_{r,k}
}
where the $\CO_{r,k}$ are KPZ dressed operators in
the Kac table, the latter being parametrized as in
\nati.
One of the couplings $\tau_{r,k}$ must
be nonzero to
``set the scale,'' i.e., some operator must provide an
infrared cutoff
on the functional integral over surfaces.
Correlation functions
for nearby perturbed theories are defined by conformal
perturbation expansions such as \backdp. The nature of
such expansions is difficult to analyze in the continuum
theory since the coefficients are
difficult to compute. Enter the matrix model.

The solution of the continuum limit of the ``$q$-matrix model''
indicates the existence of an infinite dimensional space
of coupling constants, namely the space
of real tuples $\{ t_{r,k}\}$, $1\leq r\leq q$, $0\leq k$,
with all but finitely many $t$'s $=0$
\ref\mrd{M. Douglas, ``Strings in less than
one dimension and the generalized kdv hierarchies,''
Phys. Lett. {\bf 238B}(1990)176}.
There is a certain amount of evidence
\ref\mart{See E. Martinec,
``An Introduction to 2d Gravity and Solvable
String Models,'' Rutgers preprint, hepth@xxx/ 9112019,
for a recent
review.}\
that the continuum matrix model defined by the tuple
$\bar t^{(p)}_{r,k}=\delta_{kq+r,p}$ is identical to
the theory $\CS_{p,q}$. Just as the path integral for
\minpert\ is ill-defined if $\tau_{r,k}=0$, the
$(p,q)$ string equation is singular for
$t_{r,k}-\bar t^{(p)}_{r,k}=0$. The mapping
from $t_{r,k}$ to $\tau_{r,k}$ is nontrivial and
has only been partially worked out in some special
cases
\ref\mss{G. Moore, N. Seiberg, and M. Staudacher, ``From Loops to
States in Two-Dimensional Quantum Gravity,''
Nucl. Phys. {\bf B362}(1991)665}.

The advantage of the matrix model formulation is that the string
equation and KP flow give a complete mathematical description of
the crossover phenomena for coupling constant flow between
the neighborhoods of two $(p,q)$ ``fixed points.''
Choosing some coupling,
say $x=t_{r_0,k_0}$, to set the scale, the solution of the string
equation $u(x;t_{r,k})$ will be an analytic power series in the
other couplings $t_{r,k}$. After proper identification
of the $t$'s with the $\tau$'s
this power series should correspond to
conformal perturbation theory.
At genus zero the string equations
reduce to algebraic equations for $u$.
Hence, the power series in any coupling will in general
have a finite radius of convergence.
Therefore, if we make any coupling sufficiently large
the physical description of the
system must change. Let us consider two examples of this.

Example 1. The flow from Ising to pure gravity in the
absence of a magnetic field is described by the
equation
\eqn\issteq{u^3+t u^2 =x}
Letting $x$ set the scale, for small $t$
we may identify $x$ with the
cosmological constant and $t$ with the coupling of
the thermal operator $\epsilon$. The physical
solution of \issteq\ is
determined by the branch $u=x^{1/3}$ for
$x\to \infty$. For $t<t_c\equiv (27x/4)^{1/3}$, $u(x,t)$
is a convergent power series in $t$.
In this regime the continuum description of the
theory uses the action $S=S_{\rm Liouville}+S_{\rm Ising} +
t \int e^{\xi \phi} \epsilon$.
\foot{Actually, this is only true for small $t$, and the
results of \mss\ show that there is more to
understand in this example.}
Thus, for $t<t_c$
we are perturbing the Ising fixed point by
a relevant operator.
Beyond the radius of convergence
the solution to \issteq\ must be expanded as
a convergent power series in $(x/t^3)^{1/2}$.
We are now in the neighborhood of the $(2,3)$
fixed point and in the continuum theory
we should describe this power series as a perturbation
of the pure $c=26$ Liouville
theory by a certain irrelevant operator. Thus the
action is $S=S_{\rm Liouville}+ t^{-3/2}\int \CO$
where $\CO$ is an operator in the $c=26$
Liouville theory.
Although the solution to \issteq\ is analytic in
the neighborhood of $t=t_c$ the $\sigma$-model
description of the physics changes.

Example 2. It is easy to find examples of flows
in coupling constants where there must be
a true phase transition.
Consider the string equation of
the $(2,2m-1)$ theories: $u^m+\sum_{i\geq 1} t_i u^i=x$
where we let $t_0=x$ set the scale. Consider the
graph of the function $f(u)=u^m+\sum_{i\geq 1} t_i u^i$
as a function of $u$, and denote the value of $f$ at the
local minimum with the largest value of $u$ by $h(t_i)$.
If, as we change the couplings $t_i$,
$h(t_i)$ crosses through $x$ from below
there will be a phase transition.
In this case, if we simply analytically continue
the specific heat around the branch point in complex
$t$-space $u$ will take complex values.

These examples suggest a general idea, which is
borne out by the results of this paper. There
is a strong analogy between the coordinates
$\tau_{r,k}$ on ``theory-space,'' and weighted
projective coordinates of complex manifold theory.
First, gravitationally dressed operators depend
on the Liouville zero mode only through a single
exponential factor, hence
the overall normalization of the $\tau$'s
can be changed by a shift of $\phi$. Second,
as we have remarked, the theory is singular
if all the $\tau_{r,k}=0$, reminiscent of the fact
that the origin of $\IC^{n+1}$ is in no sense
a point of $\IC P^n$. Third, in projective space regions
in which a given coordinate can be scaled to one
provide coordinate patches for the manifold.
In $\tau$-space, different coordinate patches
are defined by letting different operators
$\CO_{r,k}$ set the scale. In example 1 above
the change of expansion parameters
from $t/x^{1/3}$ to $x/t^3$ is analogous to the
change of coordinates
between two coordinate patches
of weighted projective space. In general, different
coordinate patches correspond to different phases
of the same theory.

Let us apply some of these ideas to the Sine-Gordon model.
Depending on the relative magnitudes of $m$ and
$\mu$ either
the cosmological constant or the Sine-Gordon interaction
will set the scale \nati.
If the cosmological constant sets
the scale we expect that correlators will be expressed as
power series in $m^2 \mu^{p-2}$.
For example the genus zero partition function is
$Z=\half \mu^2\log \mu+\mu^2 f(m^2 \mu^{p-2})$ where $f(z)$ has
an analytic expansion around zero.
On the other hand, if $\mu$ is small
$m$ sets the scale,
and we expand in $\mu (m^2)^{-1/(2-p)}$.
In section five below we will find that these
expectations are in accord with matrix model
calculations. We find a surprise in that there
is a phase transition, possibly analogous to
that of example 2 above, for $0<p<1$ and for $p>2$,
while the model behaves much more like example 1 above
for $1<p<2$. In section five we offer a
physical picture that describes these phase
transitions in terms of semiclassical field theory.

Finally, we conclude with a few remarks on the
relation between coupling-constant flow and
renormalization group flow. A point
where all but one $\tau$ vanishes is rather like a fixed
point of the renormalization group.
At the fixed points we have a well-defined notion
of matter central charge $c^X$ and bare matter-field
dimensions. As in flat space, the operators perturbing
away from the fixed points may be divided into relevant,
marginal, and irrelevant. Since the Liouville field $\phi$
defines the local scale, the Liouville charge $\xi$
of the KPZ dressed operator $e^{\xi \phi} \Phi_X$
is positive for relevant,
zero for marginal, and negative for irrelevant operators.
Thus, relevant operators grow in the infrared
$\phi\to +\infty$, etc. The classification of such operators
is exactly the same as in flat space since
by the KPZ formula and Seiberg's bound \nati:
$\xi={Q\over 2}
\biggl(1-\sqrt{1+{8\over Q^2}(\Delta_X-1)}\biggr)$ gives
$\xi<0$ for $\Delta^X>1$ and vice versa.
In the Sine-Gordon theory for
infinitesimal $m$ the operator
$e^{i p X/\sqrt{2}}$ is relevant for $p<2$, marginal for $p=2$
and irrelevant for $p>2$. More precisely, in flat space
the renormalization
group flow in the neighborhood of $(m,p)=(0,2)$ is given
by the flow diagram of the Kosterlitz-Thouless model
\ref\amit{D.J. Amit, Y.Y. Goldschmidt, and G. Goldin,
``Renormalisation group analysis of the phase transition in the
2D Coulomb gas, Sine-Gordon
theory and XY model,'' J. Phys. {\bf A13}(1980)585}.

Ordinary renormalization group flow is dissipative
\ref\cthrm{A. Zamolodchikov,
``Irreversibility of the flux of the
renormalization group in a 2D field theory,''
JETP Lett. {\bf 43}(1986) 730}.
In gravity one may note, phenomenologically,
that the flows by relevant perturbations of the $q$-matrix model
between two fixed points always {\it increases} $c^X$, while
$c^X_{eff}\equiv c-24 \Delta^X_{\rm min}$ always decreases.
Conversely, $c^X_{eff}$
always increases under irrelevant perturbations.
Thus, if we perturb by an irrelevant operator and we
find a phase transition we may expect that $c_{eff}^X$ has
increased provided we have
a phase transition to another surface theory.
We will show in section four
that one can perturb the Sine-Gordon
model by an irrelevant operator $p>2$ to obtain a phase
transition. The extremely interesting question of
whether this is a phase transition to a $c_{eff}^X>1$ model
remains open.

\newsec{The partition function for $m\not=0$.}

In this section we describe the result of a matrix model
calculation of a one-point function of the form
\backdp. For technical reasons it is convenient to calculate
the correlation functions of the vertex operators
$\CT_{\pm p}={\Gamma(p)\over \Gamma(-p)} V_{\pm p}$, where
$p>0$ here and hereafter. We define
the coupling constant
\eqn\chgcp{\alpha\equiv\half {\Gamma(-p)\over \Gamma(p)}m
}
Using the calculational techniques of \mpr\ and
conformal perturbation theory we have found an
explicit nonperturbative expression for
the one-point function of the cosmological constant
$\langle \CT_0\rangle_\alpha\equiv {\p\over\p \mu}\CZ$, where
\eqn\defpart{
\CZ\equiv \langle e^{\alpha \CT_p+\alpha \CT_{-p}}\rangle
}
is the partition function. The somewhat complicated
formula is given in equation $(A.6)$ of appendix A.
In particular,
defining a certain function, the ``bounce factor''
$R_q$ by:
\eqn\bncfcti{R_q\equiv \mu^{-|q|}\sqrt{2\over\pi}e^{i\pi/4}
\cos\bigl({\pi\over 2}(\half+i \mu-|q|)\bigr)\Gamma(\half-i \mu+|q|)}
we find that the amplitude
\eqn\defamp{
A_n(\mu,p) \equiv \mu^{-n p}\langle \CT_0 \CT_p^n \CT_{-p}^n\rangle}
may be expressed as a polynomial in the $R_q$ evaluated for
$q$'s at various integer multiples of $p$. See eq. $(A.1)$ for
more detail.

\subsec{String Perturbation Theory}

In order to study the partition function on a fixed topology
we must expand the nonperturbative answer $(A.1)$ in
$1/\mu$. This
may be obtained from the asymptotic expansion for the bounce factor:
\eqn\ptforr{
R_p { \buildrel \mu\to \infty \over \sim}
exp\biggl[p \tilde\psi_0+\sum_{n\geq 1} {i^n p^{n+1}\over (n+1)!}
({d\over d \mu})^n(log \mu+\tilde\psi_0)\biggr]=1+{i p^2\over \mu}
+\cdots
}
where $\tilde \psi_0$ denotes the expansion
\eqn\defpsi{
\tilde \psi_0\sim \sum_{k\geq 1}{(-1)^k B_{2k}\over 2k}(1-2^{-2k+1})
{1\over \mu^{2k}} \sim Re \Psi(\half-i \mu)-log \mu
}
$\Psi$ is the digamma function and $B_{2k}$ are Bernoulli
numbers.

Substitution of \ptforr\ into equation (A.1) for
$A_n(\mu,p)$ gives an asymptotic expansion of the form:
\eqn\asymp{
A_n(\mu,p)\sim \sum_{h\geq 0} {1\over \mu^{2n-1+2h}} A_n^h(p)
}
The very statement of KPZ scaling, namely, that the above
expansion begins at order $1\over \mu^{2n-1}$ is
somewhat miraculous from the point of view of the matrix model
and implies the existence of nontrivial combinatorial
identities on Bernoulli numbers.
Nevertheless we may extract from \ptforr\ and (A.1) the following
facts about the correlation function:

1. At each order of perturbation theory $A_n^h(p)$ is a polynomial
in $p$ of degree $4n-2+4h$.

2. $p=0$ is a zero of $A_n^h(p)$ of order $2n$.
Indeed, as $p\to 0$:
\eqn\lowen{\eqalign{
A_n(\mu,p) &\to p^{2n}\bigl({\p\over \p \mu}\bigr)^{2n-1}
Re\psi(\half - i \mu)\cr
&\sim p^{2n}\biggl[{(2n-2)!\over \mu^{2n-1}}
+{(2n)!\over 24 \mu^{2n+1}}+\cdots\biggr]\cr}
}

3. The value $p=1$ is a root of order $n$ of $A_n^h(p)$.

4. Moreover, the value $p=2$ is a root of order 1 for $A_n^h$ for $h\geq 1$,
and in general for $m$ a positive integer, $p=m$ is a root
of $A_n^h(p)$ for  $h>\half(1+ n(m-2))$.

Statement one is easily proved by examination of \ptforr.
The expansion of the term in the exponent in powers $p^a/\mu^b$
has a maximum value of $a-b$ for the term
$p^2/\mu$. In \mpr\ the amplitude
$\langle \CT_0 \prod_{i=1}^k \CT_{p_i}\rangle$ was shown to be a
polynomial in bounce factors $R_p$. It follows that at
genus $h$ the amplitude is a polynomial in $p_i$ of degree
$2k-2+4h$.
\foot{
This confirms the observation of
\ref\seishen{N. Seiberg and S. Shenker,
``A Note on Background (In)dependence,''  Rutgers preprint, hepth@xxx/ 9201017}
that at large energies the effective string coupling
in the $c=1$ model is $g_{eff}\sim {p^2\over \mu}$.}
A special case of this result is statement one.

Statement two is easily proved from the low energy
theorem in \mpr. As $p\to 0$ we have
$\CT_p\to p\CT_0=p{\p\over \p \mu}$
(except in the genus zero two-point function). Statement
\lowen\ immediately follows from the well-known value of
the specific heat
\ref\specheat{E. Brezin, V. Kazakov, Al.B. Zamolodchikov, Nucl
Phys {\bf B338}(1990)673;
P. Ginsparg and J. Zinn-Justin, Phys. Lett. {\bf B240}(1990)333;
G. Parisi, Phys. Lett. {\bf 238B}(209); D.J. Gross and N. Miljkovic,
Phys. Lett. {\bf 238B}(1990)217}.

The proof of statements three and four is sketched in
appendix B.

\subsec{The genus zero amplitude}

We now focus on
the spherical topology. The above remarks show that
$A_n^{h=0}(p)=(2n-2)!p^{2n}(1-p)^n Q_n(p)$
where $Q_n$ is a polynomial of order $n-2$ with
$Q_n(0)=1$. Finding a formula for this polynomial
has proved to be a rather difficult problem.
Explicitly expanding the formula (A.1) we find,
experimentally, the curious result
\eqn\genzer{\eqalign{
A_n^{h=0}(p)=&(2n-2)! p^{2n}(1-p)^n \prod_{i=1}^{n-2} (1-p/r_i)\cr
%
%
\langle \CT_p^n \CT_{-p}^n\rangle=&-\mu^{np-2n+2}n!p^{2n}(1-p)^n
{\Gamma(n(1-p)+n-2)\over \Gamma(n(1-p)+1)}\cr}
}
where $r_i=1+i/n$.
We emphasize that this is a phenomenological formula,
checked for $1\leq n\leq 11$. We have made many
unsuccessful attempts to prove \genzer\ for all $n$.
\foot{We thank R. Plesser for his participation in several
of these efforts.}
Two remarks might be useful to anyone else who tries:

\noindent
1. By explicit calculation, the special roots $r_i$ are {\it not}
roots of the genus one amplitude.
\foot{Thus the existence of these roots is
reminiscent of the roots of the chromatic polynomial
predicted by the Beraha conjecture
\ref\saleur{H. Saleur, ``Zeroes of Chromatic Polynomials: A new
approach to Beraha conjecture using quantum groups,'' Commun. Math. Phys.
{\bf 132}(1990)657}.}

\noindent
2. The result can be summarized as a ``Ward identity''
similar to those which have been intensively studied
recently in central New Jersey.
Specifically, defining $\CT_p=p\tilde \CT_p$ we may
use the boundary-operator Ward-identity
\ref\lpsfields{G. Moore and N. Seiberg, ``From Loops to
Fields in 2D Quantum Gravity,''  ; to appear in
Int. J. Mod. Phys. see eq. (7.4)}
\ref\mrpl{G. Moore and R. Plesser, ``Classical Scattering in
$1+1$ Dimensional String Theory,'' Yale preprint P7-92, hepth@xxx/9203060}
\def\ctt{\tilde{\CT}}
to restate the result \genzer\ as
\eqn\aswi{
\langle \ctt_0\ctt_{1+\epsilon}^n\ctt_{-1-\epsilon}^n\rangle=\langle
\ctt_{1}^n\ctt_{-1-\epsilon}^n\ctt_{n\epsilon}\rangle
}
for $0<\epsilon<1$.

In section four we will simply assume that \genzer\ holds
for all $n,p$ and explore the physical consequences.

\subsec{The special cases $p=1,2$}

Using discrete tachyon ``Ward identities'' recently
derived in \mrpl\ we can give a much more complete
description of the partition function and correlation
functions for the special cases of a Sine-Gordon background
with $p=1,2$.

At $p=1$ the dependence on $\alpha$ is
polynomial.  Correlation functions at
$\alpha\not=0$ are easily related to correlation
functions at $\alpha=0$. The general formula expressing
this relation is somewhat long, so we simply quote a typical result
\eqn\coratone{
\langle \CT_0\CT_q \CT_{-q} e^{\alpha \CT_1+\alpha\CT_{-1}}\rangle
={}_2F_1(1-q,1-q;1;\alpha^2\mu^{-1})\langle \CT_0\CT_q\CT_{-q}\rangle
}
for $q$ a positive integer.

At $p=2$ (when the matter perturbation
is formally marginal) the $\CT_2$ Ward identity of
\mrpl\ implies:
\eqn\newid{
\lim_{\epsilon\to 0^+}\langle \ctt_{-n \epsilon}
\ctt_2^n\ctt_{-2+\epsilon}^n\prod_{i=1}^k\ctt_{k_i}\prod_{j=1}^l
\ctt_{-q_j}\rangle=n!{\Gamma(Q+n)\over \Gamma(Q)}
\langle\ctt_0 \prod_{i=1}^k\ctt_{k_i}\prod_{j=1}^l
\ctt_{-q_j}\rangle}
where $k_i,q_j>0$, $\sum k_i=\sum q_j\equiv Q$, and $q_i+q_j<2$
for all pairs $i,j$. It follows that correlation functions in this
kinematic regime are given at $\alpha\not=0$ by
\eqn\correl{
\langle \prod_{i=1}^k \CT_{k_i} \prod_{j=1}^l \CT_{-q_j}
e^{\alpha \CT_2 + \alpha \CT_{-2}}\rangle=
(1-4\alpha^2)^{-Q}
\langle \prod_{i=1}^k \CT_{k_i} \prod_{j=1}^l \CT_{-q_j}\rangle
}
The correlation functions in other kinematic regimes will
not be so simply related. For example, one can show
\eqn\newidi{
\langle \CT_q \CT_{-q}e^{\alpha \CT_2 + \alpha \CT_{-2}}\rangle
=(1-4 \alpha^2)^{-q}
\bigl[1+4 \alpha^2 q(q-2)\bigr]\langle \CT_q \CT_{-q}\rangle
}
for $2<q<4$.

Using the above methods one can directly derive the specific heat:
\eqn\partattwo{
\langle \CT_0\CT_0 e^{\alpha \CT_2 + \alpha \CT_{-2}}\rangle
=\log \mu - \log(1-4 \alpha^2)
}
which is, of course, in accord with \genzer.
An amusing, and perhaps important, feature of \partattwo\
is that it is true to {\it all} orders of perturbation theory,
since as noted in section (3.1)
$p=2$ is a root of the correlation functions at genus $h\geq 1$.

We remark that the amplitudes \correl\partattwo\
exhibit an interesting
duality between the theory at
$\alpha$ and at $\tilde \alpha=1/(4\alpha)$.
This duality will be generalized below.

\newsec{Phase Diagram in the $\alpha$,$p$ plane}

According to \genzer\ the genus zero specific heat is given by
\eqn\gnzrsr{\langle \CT_0\CT_0 e^{\alpha \CT_p + \alpha\CT_{-p}}\rangle-
\langle\CT_0\CT_0\rangle= -H(p;z)
}
where $z=\mu^{p-2}\alpha^2 p^2(1-p)$ and
we have defined an analytic function
\eqn\gnseri{H(p;z)\equiv
\sum_{n\geq 1}{\Gamma(n(2-p))\over n! \Gamma(n(1-p)+1)} z^n
}
Some useful mathematical facts about the function $H(p;z)$
are collected and proved in appendix C. In particular, $H$ is
a convergent power series in $z$ for all real values of $p$
with radius of convergence $R_c(p)$ given by $(C.3)$ and
plotted in
\fig\radconv{Radius of convergence as a function of $p$.}.
On the circle of convergence the series has one or two
branch point singularities given by
\eqn\bpsing{\eqalign{
z_c(p)&=R_c(p)\qquad \quad\qquad p<1\cr
z_c^\pm(p)&=-e^{\pm i \pi p}R_c(p)\qquad 1<p<2 \cr
z_c(p)&=-R_c(p)\qquad \qquad p>2 \cr}
}
Moreover, the values of $H(p;z)$ for $|z|>|z_c|$ may be related to
values within the circle of convergence by connection
formulae similar to those for the hypergeometric function.

Using the above facts we may draw a phase diagram as shown in
\fig\phasediagram{The phase diagram in the $\alpha^2$ vs $p$
plane. The solid lines indicate lines across which there are
phase transitions.}
There are six different regions:

\noindent
I.) $0<p<1$, $0\leq \mu^{p-2}\alpha^2<R_c(p)/(p^2(1-p))$

The specific heat is a power series in $\alpha^2$
with real coefficients. There is a finite radius of
convergence with singularity of the form
$(1-z/z_c)^{1/2}$.

\noindent
II.) $1<p<2$, $0\leq \mu^{p-2}\alpha^2<R_c(p)/(p^2(p-1))$

Again we have a real power series in $\alpha^2$, but the branch
point has moved off into the complex plane, so there is no
singularity as we approach the dotted line in \phasediagram.

\noindent
III.) $2< p<\infty$, $0\leq \mu^{p-2}\alpha^2<R_c(p)/(p^2(p-1))$

There is again a finite radius of convergence
with singularity $(1-z/z_c)^{1/2}$ since the branch point has moved
back to the real axis. Of course, there is no singularity in passing
between regions $I,II,III$.

\noindent
IV.) $0<p<1$, $\mu^{p-2}\alpha^2>R_c(p)/(p^2(1-p))$

The values in this region are defined by analytic continuation around
the branch point. Explicitly,
the connection formula $(C.6)$ gives
\eqn\valfour{
\eqalign{
\langle \CT_0\CT_0 e^{\alpha \CT_p + \alpha \CT_{-p}}\rangle_{IV}= &
{1\over 2-p}\biggl[\log(\alpha^2p^2(1-p))\pm i \pi\biggr]\cr
&
+{1\over p-2}
H\biggl[p';-e^{\pm i\pi/(p-2)}{\mu\over
(\alpha^2p^2(1-p))^{1/(2-p)}}\biggr]\cr}
}
where $p'=(2p-3)/(p-2)$. We have a power series in $\mu$
with {\it complex} coefficients. The $\pm$ sign depends on
the sense in which we analytically continue around the branch
point.

\noindent
V.) $1<p<2$, $\mu^{p-2}\alpha^2>R_c(p)/(p^2(p-1))$

Now again applying the same connection formula,
\eqn\valfive{
\langle \CT_0\CT_0 e^{\alpha \CT_p + \alpha \CT_{-p}}\rangle_{V}=
{1\over 2-p}\log(\alpha^2p^2(p-1))
+{1\over p-2}H\biggl[p';-{\mu\over
(\alpha^2p^2(p-1))^{1/(2-p)}}\biggr]
}
giving a power series in $\mu$ with real coefficients.

\noindent
VI.) $2\leq p<\infty$, $\mu^{p-2}\alpha^2>R_c(p)/(p^2(p-1))$

In this region we use connection formula $(C.8)$. The result is a power
series with complex coefficients and expansion parameter
\eqn\sixpar{
(\alpha^2)^{-1/(p-1)} \mu^{-(p-2)/(p-1)}
}
the expansion is analytic neither in $\alpha$ nor in $\mu$.

The case $p=2$ requires special attention as discussed in section 3.3.

Finally, we have used the coupling $\alpha$ which is
natural from the matrix model.
Changing variables $\alpha\to m$ using \chgcp\ the
phase diagram in the $p,m$ plane looks somewhat different,
and is illustrated in
\fig\difdiag{The phase diagram using the normalization standard
for vertex operators. There are now infinitely many separated
regions where we have a transition of the type $III\to VI$.}.

\newsec{Physical interpretation}

There are phase transitions when crossing the solid lines in
\phasediagram. In this section we offer some qualitative physical
interpretations of these transitions based on semiclassical
analysis. The following considerations are only meant to be
heuristic, and it would be interesting to make them more
rigorous.

The action \pertlag\ may be written as
\eqn\semilag{
\eqalign{
S=&{1\over 4\pi \gamma^2}\biggl\{
\int d^2z \sqrt{\hat g}\bigl[{
1\over 2} (\hat\nabla \phi)^2
+e^{\phi}+{Q \gamma\over 2}\phi R(\hat g)\bigr]\cr
&+\int d^2 z \sqrt{\hat g}\bigl[
{1\over 2} (\hat\nabla X)^2+ m e^{\xi \phi}
\cos (p X)
\bigr]\biggr\}\cr}
}
where for convenience we have rescaled and shifted
\eqn\chgcon{\eqalign{
m\to 4\pi m \gamma^2(2/\mu)^{\xi/\gamma} &
\quad\qquad\phi\to \gamma \phi+\log(\mu/2)\cr
p\to p/(\gamma\sqrt{2})\qquad\quad X\to \gamma X &
\quad\qquad \xi\to \xi/\gamma\cr}
}

For $Q=2/\gamma$, $\xi=1$ we have a classical conformal
field theory. Of course, quantum effects are strong at $c=1$, but
they can be summarized by the usual KPZ/DDK renormalization
of parameters, so that $\gamma=\sqrt{2}$, $Q=\sqrt{8}$,
$\xi=1-p/2$. Working semiclassically, the precise
value of $Q$ turns out to be unimportant
so we will take the classical value $Q=2/\gamma$ for
simplicity.
The physics depends sensitively on $\xi$ so we will
leave this as a free parameter with $0<\xi<1$. The
equations of motion following from \semilag\ are then
\eqn\semeqs{\eqalign{
R(e^\phi \hat g)+1+m \xi e^{-(1-\xi)\phi} \cos p X&=0\cr
\hat \nabla^2 X+m p e^{\xi \phi} \sin p X&=0\cr}
}
We restrict attention to the sphere with background metric:
\eqn\sphere{
\hat g={|dz|^2\over (1+|z|^2)^2} \qquad\qquad \hat R=8
}
the constant solutions (``vacua'') of \semeqs\ are given
by $(X_n,\bar \phi)$ where $X_n={\pi\over p} (2n+1)$, $n\in\IZ$, and
$\bar \phi$ solves
\eqn\phicon{
8+e^{\bar \phi}-\xi m e^{\xi \bar \phi}=0}
This equation only has solutions for
\eqn\mincon{
\log(m \xi)\geq (1-\xi)\log 8-\biggl((1-\xi)\log(1-\xi)+\xi \log \xi\biggr)
}
in which case the zero-mode potential looks like
\fig\zmdpot{The zero mode potential
$V(\phi,X)=8\phi + e^\phi+m e^{\xi\phi}\cos p X$ in the
case when there are constant postive curvature solutions
to the equations of motion.}.
(When there is no solution one must introduce Lagrange
multipliers to fix the area of the surface. See \nati.)

The existence of classical solutions for $m$ larger
than a critical value explains some features of the
phase diagram of the previous section. In a phase
where a solution exists we can expand around it
and therefore we expect the partition function to
be nonsingular for $\mu\to 0$ \nati.
This is in accord with the difference between
regions $I,II$ and $IV,V$.

When the condition \mincon\ is satisfied there are
in fact {\it two} allowed constant curvatures, i.e.,
there are two solutions to  \phicon. For large
values of $m$ these are approximately
\eqn\appxsol{\eqalign{
e^{\phi_s}\sim &({8\over m\xi})^{1/\xi}\quad \ll 1\cr
e^{\phi_b}\sim &(m\xi)^{1/(1-\xi)}\gg 1\cr}
}
The existence of different constant solutions,
suggests the existence of
solutions which interpolate between
the different Sine-Gordon vacua $X_n$ and connect
$\phi_b$ to $\phi_s$ or $\phi_b$ to $\phi_b$.
We refer to such solutions as instantons, although
this is an abuse of terminology.
We may write a pair of ordinary nonlinear differential
equations for radially symmetric solutions which, in the
usual particle mechanics analogy
\ref\sidney{S. Coleman, {\it Aspects of Symmetry}, Cambridge, 1985.
}
represent motion in the inverted potential $-V(\phi,X)$
as in
\fig\inst{Particle motion of
$(X(r),\phi(r)) $ for a proposed field configuration
connecting large and small geometries.}.
We have not proved that such solutions exist
\foot{The sphere metric complicates the usual analysis.}
but will assume that such solutions and their
multi-instanton counterparts do exist, and proceed.
The geometry associated to a multi-instanton configuration
is of the form
\fig\multi{A multi-instanton configuration. Many small
instantons of scale size $r^2\sim e^{\phi_s}$ join onto
a large sphere of scale size $r^2\sim e^{\phi_b}$.
}.

We will now argue that the phase transition from
$IV\to V$ may be thought of as a nucleation of a
gas of instantons which unfreezes the classical
vacuum $(X_n,\phi_b)$. We may estimate the
action of a single instanton for large $m$ by the
contribution of the potential energy:
\eqn\insact{
e^{\phi_s}\biggl[V(\phi_s)-V(\phi_b)\biggr]\sim
{1-\xi\over \xi}8^{1/\xi}(m\xi)^{(2\xi-1)/(\xi(1-\xi))}
}
For on-shell configurations the kinetic energy should be
comparable to \insact.
On the other hand, each instanton carries an entropy factor
given by the relative areas of the large and small
geometry
\eqn\ent{
2{4 \pi e^{\phi_b}\over 4\pi e^{\phi_s}}
\sim 2 (m\xi)^{1/(\xi(1-\xi))}8^{-1/\xi}
}
Therefore, for $\xi<\half$ and sufficiently large $m$
the entropy will overwhelm the action and there will
be a phase transition from nucleation of small bubbles,
analogous to vortex unbinding in the Kosterlitz-Thouless
transition.
For $\xi>\half$ this does not happen and the geometry
remains frozen in a large sphere while the Sine-Gordon
field remains frozen in its minimum. Our calculation
is only meant to be a qualitative guide. The critical
value $\xi_c=\half$ in fact coincides with the exact
quantum result $p=1$ of the previous section, but this
is probably a coincidence.

It is interesting to compare with the phase transition
described in
\klebrev
\ref\klebsuss{I. Klebanov and L. Susskind,
Nucl. Phys.{\bf B309}(1988)175;
G. Parisi, Phys. Lett {\bf 238B}(1990)213;
D. Gross and I. Klebanov, Nucl. Phys. {\bf B354}(1991)459}.
These papers show that an infinite chain of spacetime
points sufficiently closely spaced act like a spacetime
continuum in string theory.
In our example, the chain of Sine-Gordon minima make
$X$ act like a discrete degree of freedom. The spacing
between this chain of spacetime points is
$D=2\pi (\sqrt{2}/p)$ (we now revert to the conventions of
the rest of the paper). Although we have not compactified
$X$ we may express $D$ in terms of the minimal allowed
radius of compactification. The transition at $p=1$
corresponds to $R=\sqrt{2}$, the self-dual radius, while
$p=2$ is equivalent (under duality) to $R=2\sqrt{2}$, the
Kosterlitz-Thouless radius
\ref\paul{P. Ginsparg, ``Applied Confomal Field Theory,''
in {\it Fields, Strings and Critical Phenomena}, E. Br\'ezin and
J. Zinn-Justin, eds. Elsevier 1989.}.

Let us finally consider region $III$. For $p>2$, with the
renormalized value of $\xi$ the zero-mode potential
is unbounded from below in the {\it ultraviolet}.
Nevertheless, the matrix model gives finite and
real partition functions within a finite radius of convergence
for $m$. (The matrix model definition of correlators of
irrelevant operators implicitly makes a choice of regularization
and subtraction of ultraviolet infinities.)
The most mysterious transition is the the transition
$III\to VI$.
Neither the Sine-Gordon coupling nor the
cosmological constant can set the scale. This might be a
signal of a first
order phase transition to a ``branched polymer'' phase
where the tachyon destroys the worldsheet \nati.
On the other hand, if it is a
phase transition to a surface theory,
according to the discussion of section two
it could be a transition to a $c_{eff}^X>1$ model.

\newsec{Nonperturbative Phase Transitions}

The effects of topology-change can drastically alter the
nature of gravitational phase transitions. A good example
of this is the difference in the nature of nonperturbative
and perturbative phase transitions in
the flow from the Yang-Lee edge singularity to pure gravity
\ref\dss{M.R. Douglas, N. Seiberg, and S. Shenker,
''Flow and Instability in Quantum Gravity,''
Phys. Lett. {\bf B244}(1990)381}.
Analyzing the effects of topology-change in the phase
transitions of Sine-Gordon gravity requires an understanding of
the nature of the singularities of the full nonperturbative
formula for the partition function given by equation $(A.6)$
below.

Since the partition function is expressed in terms
of a determinant:
\eqn\pertopt{
\langle \CT_0 e^{\alpha \CT_p + \alpha \CT_{-p}}\rangle
-\langle \CT_0\rangle=
i \log Det\biggl[(1+\Sigma)(1+\Sigma^*)^{-1}\biggr]}
we must
ask when the operator $\Sigma$, (defined in $(A.7)$)
ceases to be traceclass. From
$(A.7)$ we may estimate
\eqn\est{\eqalign{
\parallel \Sigma\parallel &\leq
\sum_{n\geq 1} (\mu^p \alpha^2)^n \sum_{a+b=n}
{|R_{a p}|\over (a)!^2} {|R_{b p}|\over (b!)^2}\cr
&\leq \sum_{n\geq 1}n  (C(p) \mu^p \alpha^2)^n
{|R_{n p}|\over (n!)^2}\cr}
}
where $C(p)$ is a positive constant.
For large $n$, $R_{np}$ behaves like $(n!)^p$ so for
any $0<p<2$ and small enough $|\alpha|$,
$\parallel \Sigma \parallel<1$ so the determinant exists and
defines an analytic function of $\alpha^2$ and $p$.

At $p=2$ the sum in
\est\ behaves like a power series and remains convergent.
More precisely, if we let $p=2-is$ for $s$ real we have
for large $n$:
\eqn\estii{
{|R_{n p}|\over n!^2}\leq {1\over \pi}e^{-\mu \theta(s)+\half \pi \mu}
{1\over n}\mu^{-2n} (f(s))^n
}
where
$$f(s)=4(1+s^2/4) e^{\half \pi|s|-s\theta(s)}$$
and $\theta(s)={\rm tan}^{-1}(s/2)$.
Thus if $4 C\alpha^2<1$ there is a finite interval
$-s_c<s<s_c$ along the $Re(p)=2$ axis where
the series is absolutely convergent. Thus there exists
an analytic continuation of the
determinant to the domain $Re(p)>2$, where we are perturbing
the Sine-Gordon theory by an irrelevant operator.

We have just shown that, nonperturbatively, there are
analogs of the regions $I,II,III$ of section four.
It is more difficult to see if the radius of convergence
in $|\alpha|$ will be finite.
We expect that at fixed $p$ there is a finite
radius of convergence in
$|\alpha|$. As we increase $|\alpha|$, $1+\Sigma$
probably develops a left or right
zero mode and the determinant has a logarithmic singularity,
although we have not proven this.
It is easy to show that for sufficiently large $|\alpha|$,
$\parallel \Sigma\parallel>1$ so there is no reason for
the determinant to be nonsingular. It would be interesting
to understand the singularities better and to have a
physical picture of how the phase transitions are modified
by topology-change.

\newsec{Future Directions}

There are several projects which would extend the present
work:

\noindent
1. Of course, it is important to {\it prove}
\genzer! The recent results of \mrpl\ are an important step in
this direction.

\noindent
2. We also skipped over some hard analysis in
section five, regarding the existence of instanton solutions.

\noindent
3. In section four we used the connection formulae
$(C.4),(C.6),(C.8)$. These relate different backgrounds
via the action of the discrete group $S_3$ and are thus
reminiscent of target space duality and mirror symmetry.
It would be very interesting to see if these symmetries
survive in other correlation functions.

\noindent
4. We would like to have an equally
complete understanding of the amplitudes at genus one
which are implicitly contained in $(A.1)$. These would
be most useful for understanding better the nature of
the transition $III\to VI$. If the transition is due to tachyon
dominance that should become apparent in the behavior of
the genus one amplitudes. Unfortunately, we have not
managed to recognize any special pattern in the first few amplitudes.

\noindent
5. The simple result \partattwo\ deserves to be understood better.
Naively, at $p=2$ we have a tensor product of Liouville
and Sine-Gordon
theories, but for $m\not=0$, $\cos(\sqrt{2}X)$ is not
exactly marginal so this is an illusion \amit. Indeed
\partattwo\ is not a product, but a sum of functions
of $\mu$ and $\alpha^2$.

\noindent
6. It would be
interesting to interpret the minimal models as restricted
Sine-Gordon theories
and relate the above results more directly to
the $c<1$ models.

The present paper also touches on some deeper issues.
The matrix model defines finite integrated correlators even for
the irrelevant operators. How does it choose the finite parts?
A natural guess for the underlying principle is the
$W_\infty$ symmetry of the theory. This implies a
corresponding $W_\infty$ symmetry of the continuum Liouville
$\times $ matter system. Perhaps the finite parts are chosen based on
the principle that the $W_\infty$ Ward identities must be
maintained.

A second issue is the spacetime interpretation of the Euclidean
theory. If, for example, we compactify $X$ when $\alpha=0$ then we
calculate the free energy of a string at temperature $1/R$.
What happens when $\alpha\not=0$? The Euclidean Hamiltonian
has now nontrivial $X$-dependence. Should we interpret the
calculations in terms of nonequilibrium statistical mechanics?

One may also ask about the Minkowskian analog of the above
results and the corresponding Minkowskian spacetime interpretation.
Some of the relevant issues are discussed in \mrpl.

Finally, we may ask the evident question: Do analogous phase
transitions exist in more realistic theories of gravity?

\bigskip
\centerline{\bf Acknowledgements}

We would like to thank R. Plesser and N. Seiberg for
very important conversations. We further thank N. Seiberg
for useful criticism and comments on an earlier draft of
the paper. We have also benefitted from helpful
discussions with and suggestions from
T. Banks, M. Douglas, E. Martinec,
S. Ramgoolam, N.Read, H. Saleur,
R. Shankar, S. Shatashvili, S. Shenker, and A.B. Zamolodchikov.
This work is supported by DOE grant DE-AC02-76ER03075
and by a Presidential Young Investigator Award.

\appendix{A}{Derivation of the formula for the correlation functions}

We derive the formula for $\langle \CT_0 (\CT_p \CT_{-p})^n\rangle$
by considering the $\epsilon\to 0+$ limit of the correlator
$\langle \CT_\epsilon(\CT_{p-\epsilon/n})^n (\CT_{-p})^n\rangle$
using the graphical rules of \mpr. A short calculation
yields
\eqn\simpexp{\eqalign{
A_n(\mu,p)& \equiv \mu^{-n p}\langle \CT_0 \CT_p^n \CT_{-p}^n\rangle \cr
&=i(-1)^n (n!)^2\sum_{k=1}^n {(-1)^k\over k}\sum_{a_i,b_i}
(\prod_{i=1}^k b_i^2-\prod_{i=1}^k a_i^2)\CC(a_1,\dots,b_k)
 \prod_{i=1}^k {R_{a_i p}\over (a_i)!^2} {R_{b_i p}^*\over (b_i!)^2}\cr}}
where the sum runs over all partitions $n=a_1+b_1+\cdots +a_k+b_k$
with $a_i,b_i\geq 0$
such that the denominator of
\eqn\defsea{\CC(a_1,b_1,a_2,b_2,\dots,a_k,b_k)\equiv
{1\over (a_1+b_1)(b_1+a_2)(a_2+b_2)\cdots (a_k+b_k)(b_k+a_1)}
}
is nonzero.
$R_p$ is the ``bounce factor'' of \mpr\ given by
\eqn\bncefct{R_p=\mu^{-p}\sqrt{2\over\pi}e^{i\pi/4}
\cos\bigl({\pi\over 2}(\half+i \mu-p)\bigr)\Gamma(\half-i \mu+p)}
for $p>0$.

The expression \simpexp\ can be written more succinctly
by introducing an algebra of 1-dimensional
projection operators $\CP_{a,b}$ for $a,b\in \IZ_+$, not
both zero, satisfying the relations
\eqn\algba{
\CP_{a_1,b_1} \CP_{a_2,b_2}=
{(a_1+b_1)(a_2+b_2)\over (b_1+a_2)(b_2+a_1)} \CP_{a_1,b_2}
}
For such an algebra we may write the factor $\CC$ as
\eqn\useofc{{1\over \prod_i (a_i+b_i)^2 }
Tr\biggl(\CP_{a_1,b_1}\cdots \CP_{a_k,b_k}\biggr)=
\CC(a_1,\dots b_k)
}
Such projection operators may be explicitly constructed as
operators on
a Hilbert space. Let $|z_a\rangle$, $a=0,1,\dots$ be an
ON basis. Define $|w_b\rangle=\sum_a (a+b)^{-1}|z_a\rangle$ so
that $\langle z_a|w_b\rangle =(a+b)^{-1}$. Then
$\CP_{a,b}\equiv (a+b) |z_a\rangle \langle w_b|$.

Using the projection operators $\CP_{a,b}$ the full partition
function can be nicely expressed as a determinant
of an operator $\Sigma$:
\eqn\pertopt{
\langle \CT_0 e^{\alpha \CT_p + \alpha \CT_{-p}}\rangle
-\langle \CT_0\rangle=
i \log Det\biggl[(1+\Sigma)(1+\Sigma^*)^{-1}\biggr]}
where
\eqn\defsig{
\Sigma\equiv \sum_{n\geq 1} (-\mu^p\alpha^2)^n \sum_{a+b=n}
{R_{a p}\over (a)!^2} {R_{b p}^*\over (b!)^2}{b^2\over n^2} \CP_{a,b}
}

Remarks:

1. It might be an interesting exercise to obtain the above
determinant directly from the fermion determinant in the
original free-fermion formulation of the theory.

2. Nonperturbatively the series vanishes identically at $\mu=0$.
This follows immediately since $R_pR_q^*$ is real for $\mu=0$ and
any real momenta $p,q$.

3. In a similar way one can write slightly more
complicated formulae for the three point function
$\langle \CT_0\CT_q\CT_{-q}
exp\bigl( \alpha\CT_p+\alpha \CT_{-p}\bigr)\rangle$.

\appendix{B}{Integer roots of the amplitudes}

The basic idea of the proof is very simple. To all orders of
perturbation theory the bounce factor can be replaced by
\eqn\newbounce{
\eqalign{
R_p&=\mu^{-p} e^{i\pi p/2}
{\Gamma(\half-i\mu +p)\over \Gamma(\half-i \mu)}\cr
&\sim 1+\sum_{k=1}^\infty {Q_k(p)\over \mu^k}\cr}
}
For $p=n\in \IZ_+$, $R_p$ becomes a polynomial in $1/\mu$ so
that $Q_k(p=n)=0$ for $k>n$. On the other hand, by
KPZ scaling, we know the power of $1/\mu$
for the leading term in any amplitude.
If this power exceeds the order of the
relevant polynomials then we may
prove vanishing theorems. For example, if we
put $p=m$ in \simpexp, then the expression
must be of the form $1/\mu^{n m}$ times a polynomial in
$\mu$. By KPZ scaling the genus $h$ contribution
goes like  $\sim 1/\mu^{2n-1+2h}$
so that the appropriate contribution must vanish
for $h>\half +n(m-2)/2$.

The proof that $p=1$ is an $n^{th}$ order zero is much
more tedious but uses the same idea. Having proved
$p=1$ is a root we take a derivative with respect to
$p$ of \simpexp. Plugging in $p=1$ and using properties
of gamma and polygamma functions we show that ${\p\over \p p}A_n$
has an expansion in $1/\mu$ terminating at $1/\mu^{n}$.
The proof then proceeds inductively and the inductive step
fails when we consider $({\p\over \p p})^n A_n$.

We may note parenthetically that by the above reasoning {\it any}
$c=1$ amplitude with integral external momenta
vanishes at sufficiently
large orders of perturbation theory.
This supports the general idea
that special tachyons are associated
with topological field theory.

\appendix{C}{Properties of the function $H(p;z)$}

In this appendix we prove some useful facts about
the function
\eqn\gnseria{\eqalign{
H(p;z) &\equiv
\sum_{n\geq 1}{\Gamma(n(2-p))\over n! \Gamma(n(1-p)+1)} z^n\cr
&={1\over \pi}\sum_{n\geq 1}{\Gamma(n(2-p))\Gamma(n(p-1))\over n! }
\sin(n \pi p)(-z)^n\cr
&=-\sum_{n\geq 1}{\Gamma(n(p-1))\over n! \Gamma(n(p-2)+1)} (-z)^n\cr}
}

The ratio of gamma functions behaves at large $n$ like
\eqn\ratgam{\eqalign{
n^{-3/2} (exp\bigl[(2-p)log(2-p)-(1-p)log(1-p)\bigr])^n & \qquad 0<p<1\cr
n^{-3/2} (-1)^n
sin(n\pi p)(exp\bigl[(2-p)log(2-p)+(p-1)log(p-1)\bigr])^n & \qquad 1<p<2\cr
n^{-3/2} (-1)^n
(exp\bigl[(p-2)log(p-2)-(p-1)log(p-1)\bigr])^n & \qquad 2<p<\infty\cr}
}
showing that the series defining $H(p;z)$ converges absolutely
for
\eqn\radconv{
|z|<R_c(p)=exp\biggl[(p-2) \log|p-2| - (p-1) \log|p-1|
\biggr]
}
Note that from comparing the first and third lines we have
the first connection formula:
\eqn\confrmi{H(p;z)=-H(3-p;-z)
}

We can define analytic continuations of the function $H(p;z)$
using various integral representations.

Our first integral representation is the Mellin-Barnes representation
\eqn\intrepi{
{1\over 2\pi i}\int_{-i \infty}^{i\infty}
{\Gamma(s(2-p))\over \Gamma(s(1-p)+1)}
\Gamma(-s) (-z)^s ds
}
where $z\notin\IR^+$ and we use the standard branch of
the logarithm. The integral over $s$
converges absolutely for all such $z$ if $p<1$ and converges
for $|arg(z)|\geq \pi(p-1)$ for $1<p<2$.
For $|z|<|z_c(p)|$ we can close the $s$ integral in the
right half-plane to obtain the series in the first line of
\gnseria.
If $|z|>|z_c(p)|$ then we can close
in the left half-plane, thus proving the connection formula
\eqn\confrmii{
H(p;z)={1\over p-2} \log(-z) + {1\over 2-p} H(p';-(-z)^{-1/(2-p)})
}
for $p'-2={1\over p-2}$.

Our second integral representation is the Mellin-Barnes integral
\eqn\intrepii{
{-1\over 2\pi i}\int_{-i \infty}^{i\infty}
{\Gamma(s(p-1))\over \Gamma(s(p-2)+1)}\Gamma(-s) z^s ds
}
where $z\notin \IR^-$.
For $p>2$ this converges absolutely for all such $z$ and defines
an analytic continuation of $H$. For $1<p<2$ the integral
converges absolutely for $|arg(z)|\leq \pi(p-1)$.
By closing into the right
half-plane we obtain the series in the third line of
\gnseria\ and by closing in the left half-plane we obtain
the third connection formula:
\eqn\confrmiii{
H(p;z)= {-1\over 1-p} \log z + {1\over p-1} H(p'; z^{-1/(p-1)})
}
where $p'-1={1\over p-1}$.

Our third integral representation is derived from the
middle series in \gnseria\ using the integral representation of
the Beta function. The result is
\eqn\intrepiii{
z{\p\over \p z} H(p;z)=-{z \sin \pi p\over
\pi}\int_0^1 dt {t^{1-p}(1-t)^{p-2}\over
\bigl(1+z e^{i \pi p} t^{2-p}(1-t)^{p-1}\bigr)
\bigl(1+z e^{-i \pi p} t^{2-p}(1-t)^{p-1}\bigr)}
}
This integral always converges at the endpoints $t=0,1$ for
$1<p<2$ and defines an analytic continuation in $z$ for these
values of $p$. The existence of singularities can be examined
by looking for pinching of the contour. In this way it is easy
to check that
there are no singularities as $z$ increases from zero
to infinity
through real values.

In general $H(p;z)$ does not seem to be expressible in
terms of standard special functions, although at some
special values we can write $H$ more explicitly:
\eqn\specvles{\eqalign{
H(0;z)&=-\log \biggl({1+\sqrt{1-4 z}\over 2}\biggr)\cr
H(\half;z)&=\pi \int_0^z\biggl[{}_2F_1({5\over 6},{7\over 6};{3\over 2};
{27 t^2\over 4})-1\biggr]dt \cr
&+{2\over 3}\pi \int_0^z\biggl[{}_2F_1({1\over 3},{2\over 3};{1\over 2};
{27 t^2\over 4})-1\biggr]{dt\over t}\cr
H(1;z)&=-\log(1-z)\cr
H({3\over 2};z)&=2\log\biggl[\sqrt{1+{z^2\over 4}}+{z\over 2}\biggr]\cr
H(2;z)&=\log(1+z)\cr}
}
When $p$ is rational, $H(p;z)$ can be written in terms of generalized
hypergeometric functions ${}_aF_b$. In general for fixed $p$ the
branch point singularity in $z$ is a square root singularity, except
at $p=1,2$ where we have a logarithmic singularity.

Finally we note that the set of transformations of the $(p,z)$
plane defined in \confrmi\confrmii\confrmiii\ defines an
action of the permutation group $S_3$ on this plane.

\listrefs
\listfigs

\bye